# Determination of absolute densities of B, Al, Ga and Si atoms in non-equilibrium plasmas from relative intensities in resonance multiplets distorted by reabsorption

B. P. Lavrov, M. S. Ryazanov

*Faculty of Physics, St.-Petersburg State University, St.-Petersburg, 195904, Russia*
E-mail: lavrov@pobox.spbu.ru

The approach proposed recently as a spectroscopic tecnique of a boron density determination in microwave plasma has been expanded for B, Al, Ga, Si atoms and generalized. The method is based on measurements of relative intensities of pairs of spectral lines belonging to resonance multiplets having common upper level. It is shown that when the reabsorption of the lines within plasma is sufficient, then the intensity ratios depend on the gas temperature $T$ and the product $NL$ of total density of the ground state atoms $N$ and a length of a plasma column $L$. Thus in the case of the B, Al, Ga doublets the temperature $T$ should be known. The resonance multiplet of Si (three independent intensity ratios) provides an opportunity to obtain both $T$ and $N$ and even to organize internal check of the applicability of the proposed method to certain plasma conditions. The ratios were calculated for resonance lines of B, Al, Ga and Si under $T = 300 \div 2400$ K and $NL = 10^7 \div 10^{15}$ cm$^{-3}\cdot$m. The results of numerical calculations and ranges of applicability of the method are discussed.

It is well-known that the ratio of total (integrated over wave numbers) emission intensities of multiplet components having common upper level depends on partial densities of lower levels when reabsorption is sufficient. But it was not used for determination of atomic densities in plasma, until recently this effect has been proposed and used for spectroscopic determination of atomic boron density in microwave plasma [1]. The goal of present work consists in generalization of this approach for another elements and studies of its ranges of applicability.

Observable intensity ratios of multiplet components may be linked with total density of atoms $N$ and the gas temperature $T$ (in the case of resonance doublet of boron the dependence on $T$ has been neglected in [1]) under following assumptions:

1) plasma is homogeneous;

2) a population of the ground state (spread among sublevels of a multiplet structure) is much higher than that of all excited states;

3) the sublevels are populated according to the Boltzmann law;

4) the profiles of the emission and absorption lines are identical, being determined by Doppler broadening with Gaussian shape (corresponding to Maxwellian velocity distributions of excited and ground state atoms with the same value of $T$).

Then the measured total line intensity $I_{ki}$ (a total flux of photons coming out of the light source) corresponding to the $k \to i$ transition may be written as [2,3]

$$I_{ki} = A_{ki} N_k L S(\varkappa_{ik}^0 L), \qquad (1)$$

where

$$\varkappa_{ik}^0 = \sqrt{\frac{\mu}{2\pi RT}} \frac{1}{8\pi \nu_{ki}^3} \frac{g_k}{g_i} A_{ki} N_i,$$

$A_{ki}$ is Einstein coefficient for the $k \to i$ transition due to spontaneous emission, $N_k$, $N_i$ are population densities of upper and lower levels, $S$ is Ladenburg-Levi function [2], $\mu$ — atomic mass (g/mol), $R$ — molar gas constant, $\nu_{ki}$ — wave numbers (cm$^{-1}$), $g_k$, $g_i$ are degeneracities of the levels.

According to the Boltzmann law

$$N_i = N \frac{g_i e^{-\frac{E_i}{kT}}}{Z(T)}, \qquad (2)$$

where statistical sum
$$Z(T) = \sum_i g_i e^{-\frac{E_i}{kT}},$$

$N$ is the total density of the ground state (close to the total density of atoms in the plasma according to the assumption 2), and $E_i$ is an energy of $i$-th sublevel of a multiplet structure.

Finally, intensity ratio is
$$R_{knm}(T, NL) = \frac{I_{kn}}{I_{km}} = \frac{A_{kn}}{A_{km}} \frac{S(C_{kn} f_n(T) \cdot NL)}{S(C_{km} f_m(T) \cdot NL)}, \qquad (3)$$

where
$$C_{ki} = \sqrt{\frac{\mu}{2\pi R}} \frac{g_k A_{ki}}{8\pi \nu_{ki}^3}, \quad f_i(T) = \frac{e^{-\frac{E_i}{kT}}}{\sqrt{T} Z(T)}.$$

One may see that an observable intensity ratio is connected with total density $N$, the length of plasma column $L$ and gas temperature $T$. Thus, measurements of $I_{kn}/I_{km}$, $T$ and $L$ may be used for spectroscopic determination of the density $N$ in favourable conditions, when $I_{kn}/I_{km}$ has enough sensitivity to $N \cdot L$ (see e.g. [1]).

The dependencies of the intensity ratios on the optical length $NL$ for components of resonance multiplets of B, Al, Ga and Si atoms were calculated for $T = 300 \div 2400$ K, that covers typical conditions in low-temperature plasmas. Numerical data concerning spectral lines under the study are listed in table 1. We used $NL$ in cm$^{-3}$·m, then numerical value of $NL$ is equal to the particle density for the length of plasma column $L = 1$ m typical for discharge tubes. The results of the calculations are shown on figures 1—5.

One may see that the range of applicability of the method depends significantly on the value of the temperature. For example, for atomic aluminium at $T = 300$ K limits of measurable $NL$ are $(1 \div 6) \cdot 10^{10}$ cm$^{-3}$·m, while at $T = 2400$ K the product $NL$ may be obtained in the range $(0.7 \div 100) \cdot 10^{10}$ cm$^{-3}$·m (for the 5% precision of the relative intensity measurements).

In the case of B the intensity ratio becomes insensitive to temperature $T$ for high enough optical thicknesses $NL$. In contrast the same plots for Al, Ga and Si show rather strong dependencies on $T$ even for high values of optical thickness $NL$ what may be used for an estimation of $T$.

The peculiarity and main advantage of the proposed method is the opportunity to determinate **absolute** particle densities from **relative** intensity measurements, and almost without calibrations for narrow multiplets. The necessity of independent temperature determination is its main problem.

The resonance multiplet of Si presents an interesting opportunity to organize the internal check of the method by use of three independent intensity ratios (one for $2s^2 3p4s\ ^3P_2^o \to 2s^2 3p^2\ ^3P_{1,2}$ doublet and two for $2s^2 3p4s\ ^3P_1^o \to 2s^2 3p^2\ ^3P_{0,1,2}$ triplet). Moreover, one may use the triplet for simultaneous determination of $NL$ and $T$ by solution of corresponding system of nonlinear equations and the doublet for the independent check.

### Acknowledgement

Authors are thankful for financial support from the Russian Foundation for support of Basic Research (RFBR), Grant No 03-03-32805.

### References

[1] B. P. Lavrov, M. Osiac, A. V. Pipa, J. Röpcke, Plasma Sources Sci. Technol, v. 12, p. 576-589 (2003).
[2] R. Ladenburg, S. Levy, ZS. f. Phys. VI, 189 (1930).
[3] Spectroscopy of Gas Discharge Plasmas, edited by S. E. Frisch (Nauka, Leningrad, 1970) (in Russian).

Table 1.

| Atom | $\lambda_{ki}$, nm | $E_i$, cm$^{-1}$ | $g_i$ | $g_k$ | $A_{ki}$, $10^8$ s$^{-1}$ |
|---|---|---|---|---|---|
| B  | 249.75243 | 0       | 2 | 2 | 0.864 |
|    | 249.84762 | 15.254  | 4 | 2 | 1.72  |
| Al | 394.51224 | 0       | 2 | 2 | 0.527 |
|    | 396.27410 | 112.061 | 4 | 2 | 1.04  |
| Ga | 417.206   | 826     | 4 | 2 | 1.06  |
|    | 403.298   | 0       | 2 | 2 | 0.54  |
| Si | 250.76522 | 77.115  | 3 | 5 | 0.627 |
|    | 251.68696 | 223.157 | 5 | 5 | 1.86  |
|    | 251.50725 | 0       | 1 | 3 | 1.829 |
|    | 251.99600 | 77.115  | 3 | 3 | 0.618 |
|    | 252.92682 | 223.157 | 5 | 3 | 1.02  |

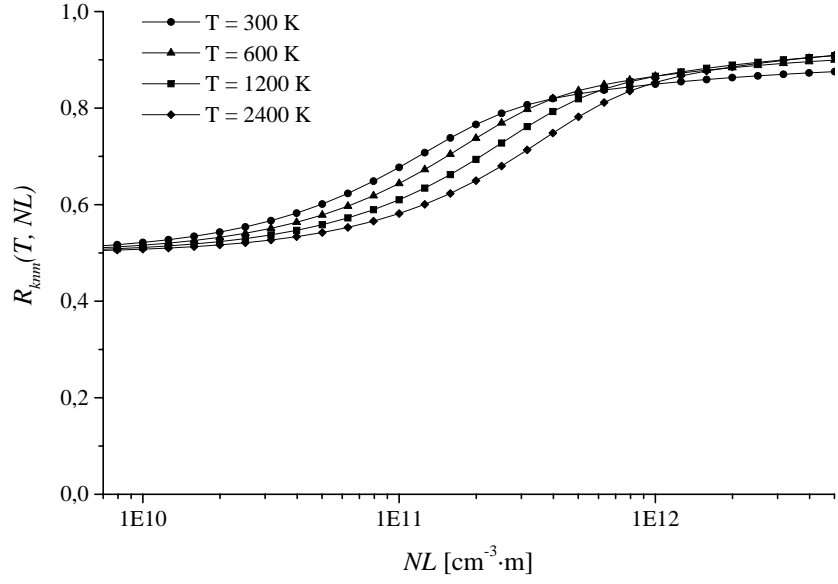

Figure 1: The intensity ratio $R_{knm}(T, NL)$ for resonance doublet of B atoms calculated as a function of the optical thickness $NL$ for various gas temperatures $T$.

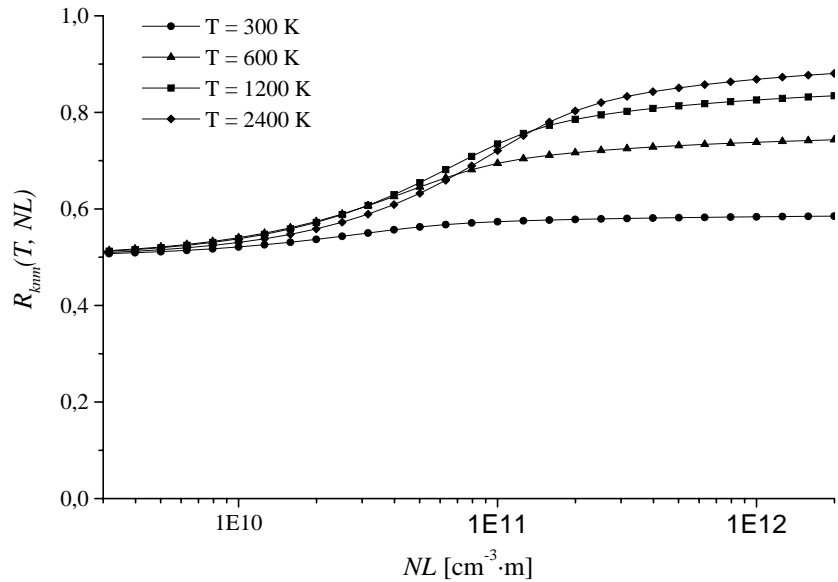

Figure 2: The intensity ratio $R_{knm}(T, NL)$ for resonance doublet of Al atoms calculated as a function of the optical thickness $NL$ for various gas temperatures $T$.

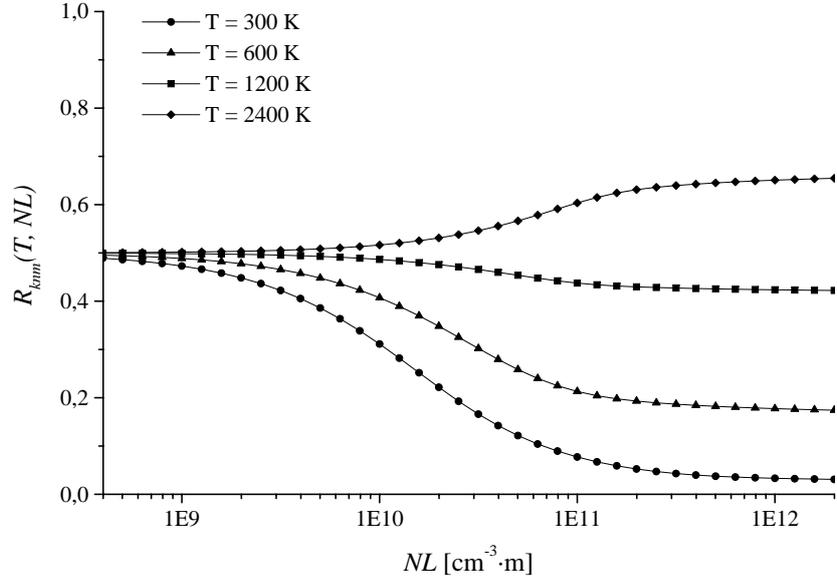

Figure 3: The intensity ratio $R_{knm}(T, NL)$ for resonance doublet of Ga atoms calculated as a function of the optical thickness $NL$ for various gas temperatures $T$.

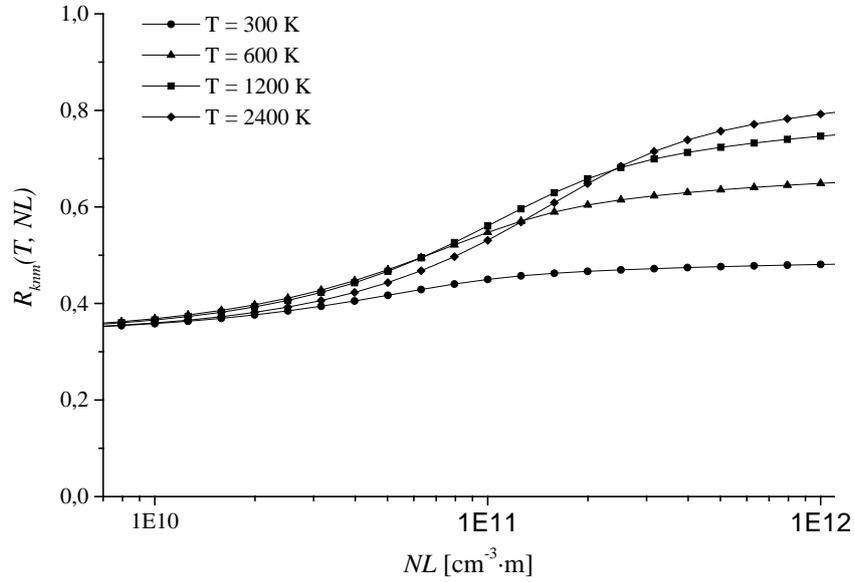

Figure 4: The intensity ratio $R_{knm}(T, NL)$ of 251.68696 and 250.76522 nm spectral lines of Si atoms calculated as a function of the optical thickness $NL$ for various gas temperatures $T$.

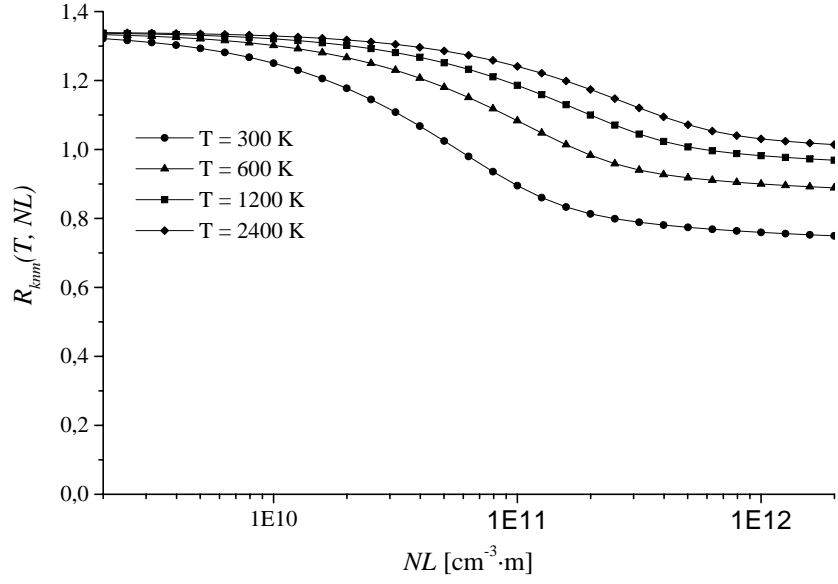

Figure 5: The intensity ratio $R_{knm}(T, NL)$ of 251.50725 and 251.99600 nm spectral lines of Si atoms calculated as a function of the optical thickness $NL$ for various gas temperatures $T$.

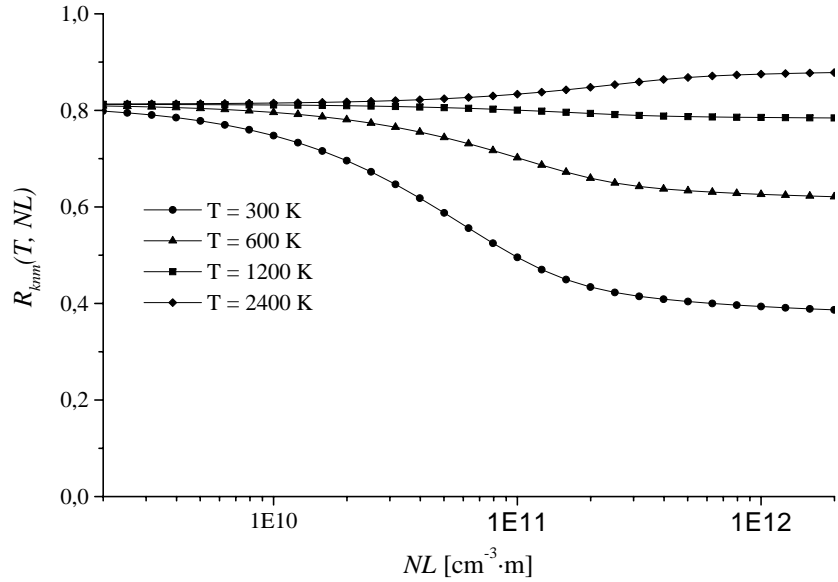

Figure 6: The intensity ratio $R_{knm}(T, NL)$ of 251.50725 and 252.92682 nm spectral lines of Si atoms calculated as a function of the optical thickness $NL$ for various gas temperatures $T$.